\documentclass[acmlarge]{acmart}
\AtBeginDocument{%
  \providecommand\BibTeX{{%
    \normalfont B\kern-0.5em{\scshape i\kern-0.25em b}\kern-0.8em\TeX}}}

\setcopyright{acmcopyright}
\acmDOI{XXXXXXX.XXXXXXX}

\begin{document}

\title{Should Policymakers be Involved? Understanding the Opinions and Needs for Independent Food Delivery Platforms in the United States regarding Public Policy}


\author{Yuhan Liu}
\affiliation{%
  \institution{Princeton University}
   \country{USA}
  }
\email{yl8744@princeton.edu}

\author{Amna Liaqat}
\affiliation{%
  \institution{Princeton University}
   \country{USA}
  }
\email{al0910@princeton.edu}

\author{Andrés Monroy-Hernández}
\affiliation{%
  \institution{Princeton University}
   \country{USA}
  }
\email{andresmh@princeton.edu}

\renewcommand{\shortauthors}{Liu et al.}

\begin{abstract}
Mainstream food delivery platforms, like DoorDash and Uber Eats, have been the locus of fierce policy debates about their unfair business and labor practices. At the same time, hundreds of independent food delivery services provide alternative opportunities to many communities across the U.S. We surveyed operators of independent food delivery platforms to learn about their perception of the role of public policy. We found conflicting opinions on whether and how policy should interact with their businesses, ranging from not wanting policymakers to interfere to articulating specific policies that would curtail mainstream platforms' business practices. 
We provide insights for technologists and policymakers interested in the sociotechnical challenges of local marketplaces.
\end{abstract}



\maketitle

\section{background}
In this study, we investigate the needs and challenges of indie \footnote{We borrow the term ``indie'' from the gaming and film industry to refer to platforms that are not one of the large mainstream ones, i.e., Uber Eats, Doordash and Grubhub} food delivery platforms. 
Indie platforms typically focus their efforts on a small and localized scale, having more human intervention in their business operations than mainstream platforms. This helps them differentiate themselves in a competitive market~\cite{atkinsonMoreJobFood2021, schneiderExitCommunityStrategies2020, nosh2023}. Our previous studies have discussed the sociotechnical limitations of indie platforms but have not covered how platform operators see the role of public policy in their business~\cite{indie2023cscw, nosh2023}. In our ongoing project, we investigate the opinions and needs of indie platform operators' perspectives regarding public policy and how policy can support or hinder their business operations.  This paper presents the answers to policy questions on a survey of 24 operators of independent food delivery platforms. We identify opportunities for HCI researchers to design public policy and technology simultaneously for the food delivery market and other marketplaces.

\section{Survey and results}
We launched a survey in collaboration with the RMDA (Restaurant Marketing Delivery Association), a nonprofit organization of independent food delivery services. We distributed our survey through the RMDA's member mailing list. In the survey, we asked multiple-choice, and open-ended questions about the challenges indie platforms are facing and what kind of support they need to tackle the challenges from the public policy perspective. In this workshop paper, we share the results of two of the questions that concern public policy:

\begin{enumerate}
    \item  Which of the following challenges is [name of their platform] facing? Select all that apply.
        \begin{itemize}
            \item Shortage of restaurants
            \item Too many restaurants want to sign up
            \item Shortage of couriers, e.g., drivers
            \item Oversupply of couriers, e.g., drivers
            \item Shortage of employees
            \item Too many delivery services and not enough customers
            \item Low volume of orders
            \item Shortage of customers
            \item Lack of funding
            \item Others (please specify)
        \end{itemize}
    \item (Open-ended question) How could policymakers help [the name of their platform] tackle the challenges it faces, if at all?
\end{enumerate}

Twenty-four platform operators responded. We conducted a thematic analysis of their responses to the open-ended questions, and we report on the results in the following section.

\section{Results}
The most common challenge platforms face is a lack of couriers (17 out of 24), followed by a low volume of orders (10 out of 24), a lack of funding (8 out of 24), a shortage of restaurants (7 out of 24), too many delivery services and not enough customers (5 out of 24), a shortage of customers (4 out of 24), and too many restaurants want to sign up (2 out of 24). One platform faced the challenge of a shortage of employees. No platform indicated they have an oversupply of couriers. 

We found that more than half of the survey respondents, specifically 13 out of 24, have specific suggestions on what should be changed currently regarding policy, while 2 out of 24 platforms believe that the current situation does not need to be changed. Seven replied with ``I'm not sure,''  ``I have no idea'' or similar. Two indie platform operators mentioned that their platforms are on a small scale, and they do not see how  public policy would impact their business. 

\subsection{Skepticism About the Role of Public Policy}
Several indie platform operators mentioned not wanting policymakers to interfere with their business. These respondents seem to equate policymakers to politicians and believe that policymakers do not understand their needs since they may not know how indie platforms operate. For instance:
\begin{quote}
    ``The best thing they can do to help is stay out of my way. We keep our rates fair and provide a great service. We don't need a politician to `fix' what isn't broken.'' (R4)
\end{quote}
\begin{quote}
    ``Like all businesses, we know our costs and charge what we need to be profitable. Legislators do not know our margins.'' (R18)
\end{quote}
Another viewpoint is that public policy exclusively benefits mainstream platforms, and indie platforms will not be affected due to the small scale they are operating. Respondents stated:
\begin{quote}
    ``Polices only help companies like Doordash or the like.'' (R2)
\end{quote}
\begin{quote}
    ``We aren't big enough to have this affect us.'' (R10)
\end{quote}

\subsection{Transparency, Lower Taxation, and Consent}
On the other hand, some indie platform operators articulated ways public policy could address various problems. We organized their needs into four categories: transparency, tax rates, restaurant consent, and others. 

For transparency, three respondents advocate that the platforms in the food delivery market should disclose information such as their commission rates from restaurants, the compensation they pay for couriers, and price markups on restaurant menus. For instance: 
\begin{quote}
    ``I think the biggest help would create a policy that provides complete transparency of pricing, and commission the service is taking from the restaurants. Not being able to combine the Sales Tax and Service Fee in the same line item.'' (R6)
\end{quote}

For tax rates, three respondents expressed their need for lowering tax rates, especially delivery fee sales tax, due to the current difficulty they have in making a profit: 
\begin{quote}
    ``Stop taxing us on delivery fees as all the little fees we pay makes it extremely hard to make any profit let alone continue to pay for all the expenses to be able to grow and maintain.'' (R24) 
\end{quote}

Another common need we found through the survey is restaurant consent. Preliminary studies have indicated that mainstream platforms list restaurants without their authorization.\cite{nosh2023}. This poses a problem for restaurants but also for indie platforms because they may have negotiated some deal with local restaurants for being exclusively listed on their platform. As pointed out by one of our survey respondents:
\begin{quote}
    ``We have exclusive deals set up with small locally owned restaurants and they constantly pop up on the national platforms. The big companies make it next to impossible for them to get themselves removed as well.'' (R11)
\end{quote}

Other needs we identified that were less commonly mentioned included access to parking, loans, and insurance. One respondent implied the inconvenience of getting parking spots for general delivery services in crowded vehicle areas in LA:
\begin{quote}
    ``Regulation on parking LA County / LA Metro is a crowdy vehicle area and lacks spaces for the park at the time of deliveries or pickups including harassing parking enforcement departments and the local police department, building structures with high fees of parking tickets without validation for RDS.'' (R23)
\end{quote}
Another respondent mentioned how indie platforms could benefit from an easier way to get SBA \footnote{Loans guaranteed by the Small Business Administration, see https://www.sba.gov/funding-programs/loans} loans. R3 suggested opening up regulations on alcohol delivery. 
\section{Discussion}

We organize this section with three main takeaways. First, from analyzing the responses to the two questions, we learned that there is no direct connection between the challenges indie platforms face and their expectations of public policy, which indicates the necessity of understanding their needs. Second, our findings showed conflicting opinions regarding public policy in the food delivery market. Some stakeholders think policymakers should not be involved in the food delivery market, while others are open to changes with specific suggestions. We bring up the question to the HCI community about how to reconcile these controversial opinions. In our third insight, we discuss design implications inspired by the study and hope to generalize our findings to other marketplaces and inspire future HCI research. 

\subsection{Necessity of the Understanding the Need of Indie Platforms}
In the previous section, we reported the statistical results of the questions about challenges faced by indie platforms and the support they needed from policymakers. We do not see a connection between the obstacles encountered by indie platforms and the specific public policies they wish to see put into effect. One explanation is that indie platforms could benefit if mainstream platforms were restricted to some extent, and they depend on policymakers to fulfill that. We illustrate our research findings here and hope it could provide insights for future related HCI research. 

\subsection{How to Reconcile Controversial Opinions in Designing of Public Policy}
In this study, we saw conflicting opinions about whether policymakers should be involved in the market. Previous studies have pointed out that there are different beliefs on whether the market should adjust by itself without the government's invention~\cite{timmer1989food, little1982economic, braverman1986rural}. We noticed the same debate in the food delivery market. At the Policy and Tech workshop, we intend to brainstorm policy solutions for reconciling the conflicted ideas presented by indie platform owners.

\subsection{Easily-Adapted Technology to Reflect Diverse Opinions}
As previous studies proposed, a potential technical solution to the problem indie platforms face is building public-interest software~\cite{nosh2023, indie2023cscw}. Due to the diverse opinions regarding public policy, the needs of stakeholders could be very different. So we suggest that when designing public-interest technologies, it should be easy to adapt according to different policy requirements. One potential way is to utilize and deploy numerous small, independent microservices instead of bundling all functionalities into a single monolithic application. In this way, fast and agile changes in the system could be made according to the policy changes ~\cite{chen2017monolith}.  Moreover, as we mentioned above, the phenomena we observed in the food delivery market may not be a single case, similar design insights about policy and technology could also be applicable in other marketplaces. 
\section{Conclusion}
In this paper, we shared the survey results for the challenges indie platforms face and their opinions and needs toward public policy. We observed divided opinions about whether to involve policymakers in the food delivery market: interest to change the status quo with public policy and the belief that public policy should make no invention to the market.
In addition, we discussed the limitation of our paper and raise the question to the HCI community about how to integrate two-sided opinions into the design of public technology. Lastly, we proposed design implications about easily-adapted technologies to reflect diverse opinions.

\bibliographystyle{ACM-Reference-Format}
\bibliography{bib}

\appendix

\end{document}